\documentclass[letterpaper]{article} % DO NOT CHANGE THIS
\usepackage{aaai25}  % DO NOT CHANGE THIS
\usepackage{times}  % DO NOT CHANGE THIS
\usepackage{helvet}  % DO NOT CHANGE THIS
\usepackage{courier}  % DO NOT CHANGE THIS
\usepackage[hyphens]{url}  % DO NOT CHANGE THIS
\usepackage{graphicx} % DO NOT CHANGE THIS
\usepackage{natbib}  % DO NOT CHANGE THIS AND DO NOT ADD ANY OPTIONS TO IT
\usepackage{caption} % DO NOT CHANGE THIS AND DO NOT ADD ANY OPTIONS TO IT
\usepackage{algorithm}
\usepackage{algorithmic}

\usepackage{newfloat}
\usepackage{listings}
\DeclareCaptionStyle{ruled}{labelfont=normalfont,labelsep=colon,strut=off} % DO NOT CHANGE THIS
\floatstyle{ruled}
\newfloat{listing}{tb}{lst}{}
\floatname{listing}{Listing}
\usepackage{algorithm}
\usepackage{algorithmic}
\usepackage{pdfpages}
\usepackage{subfigure}
\usepackage{caption}
\usepackage{subcaption}

\usepackage{xcolor}

\usepackage{newfloat}
\usepackage{listings}
\DeclareCaptionStyle{ruled}{labelfont=normalfont,labelsep=colon,strut=off} % DO NOT CHANGE THIS
\floatstyle{ruled}
\newfloat{listing}{tb}{lst}{}
\floatname{listing}{Listing}
\usepackage{xspace}

\newcommand*{\weareai}{\emph{We are AI}\xspace}

\title{We Are AI: Taking Control of Technology}

\author{
    Julia Stoyanovich\textsuperscript{\rm 1},
    Armanda Lewis\textsuperscript{\rm 1},
    Eric Corbett\textsuperscript{\rm 2}, 
    Lucius E.J. Bynum\textsuperscript{\rm 1},
    Lucas Rosenblatt\textsuperscript{\rm 1},
    Falaah Arif Khan\textsuperscript{\rm 1}
}
\affiliations{
    \textsuperscript{\rm 1}New York University, New York, NY, USA\\
     \textsuperscript{\rm 2}Google Research, New York, NY, USA\\
    \{stoyanovich,al861,lucius,lucas.rosenblatt,fa2161\}@nyu.edu,
    ecorbett@google.com
}

\begin{document}

\maketitle

\begin{abstract}

Responsible AI (RAI) is the science and practice of ensuring the design, development, use, and oversight of AI are socially sustainable---benefiting diverse stakeholders hile controlling the risks. Achieving this goal requires active engagement and participation from the broader public. This paper introduces ``We are AI: Taking Control of Technology,'' a public education course that brings the topics of AI and RAI to the general audience in a peer-learning setting.  

We outline the goals behind the course's development, discuss the multi-year iterative process that shaped its creation, and summarize its content. We also discuss two offerings of \weareai to an active and engaged group of librarians and professional staff at New York University, highlighting successes and areas for improvement. The course materials, including a multilingual comic book series by the same name, are publicly available and can be used independently. By sharing our experience in creating and teaching \weareai, we aim to introduce these resources to the community of AI educators, researchers, and practitioners, supporting their public education efforts.

\end{abstract}

\section{Introduction}
\label{sec:intro}
The growing prevalence of AI has sparked widespread discussions on meaningfully engaging diverse audiences this technology.  Research highlights a significant gap in participation from various stakeholders in the creation, deployment, and ultimate use of AI systems.  Facilitating accessible entry points for non-technical stakeholders to understand the basic principles of AI technology is essential for giving the public real agency with respect to AI. 

Responsible AI (RAI) has emerged as a key field to study the consequences and advantages of AI while mitigating its detrimental effects from legal, socio-cultural, and technical perspectives ~\cite{ali_discrimination_2019, Barocas2016, kitchin_towards_2014, DBLP:conf/edbt/StoyanovichYJ18}. While RAI integrates multiple disciplines, its education remains largely focused on those with technical or organizational expertise~\cite{dominguez_figaredo_responsible_2023}.  This highlights the need to engage non-experts and distill RAI insights for broader audiences affected by AI. 

This article presents our experience  with \emph{We Are AI: Taking Control of Technology}, a project aimed at creating a sustainable, scalable model for diverse audiences to learn about AI and RAI.  We describe the motivation and theoretical underpinnings behind \weareai, detail our design approach, and report on three iterations of the course, demonstrating positive learning outcomes and strong participant engagement among non-expert stakeholders.  

\begin{figure}[t!]
\centering
\includegraphics[width=.45\textwidth]{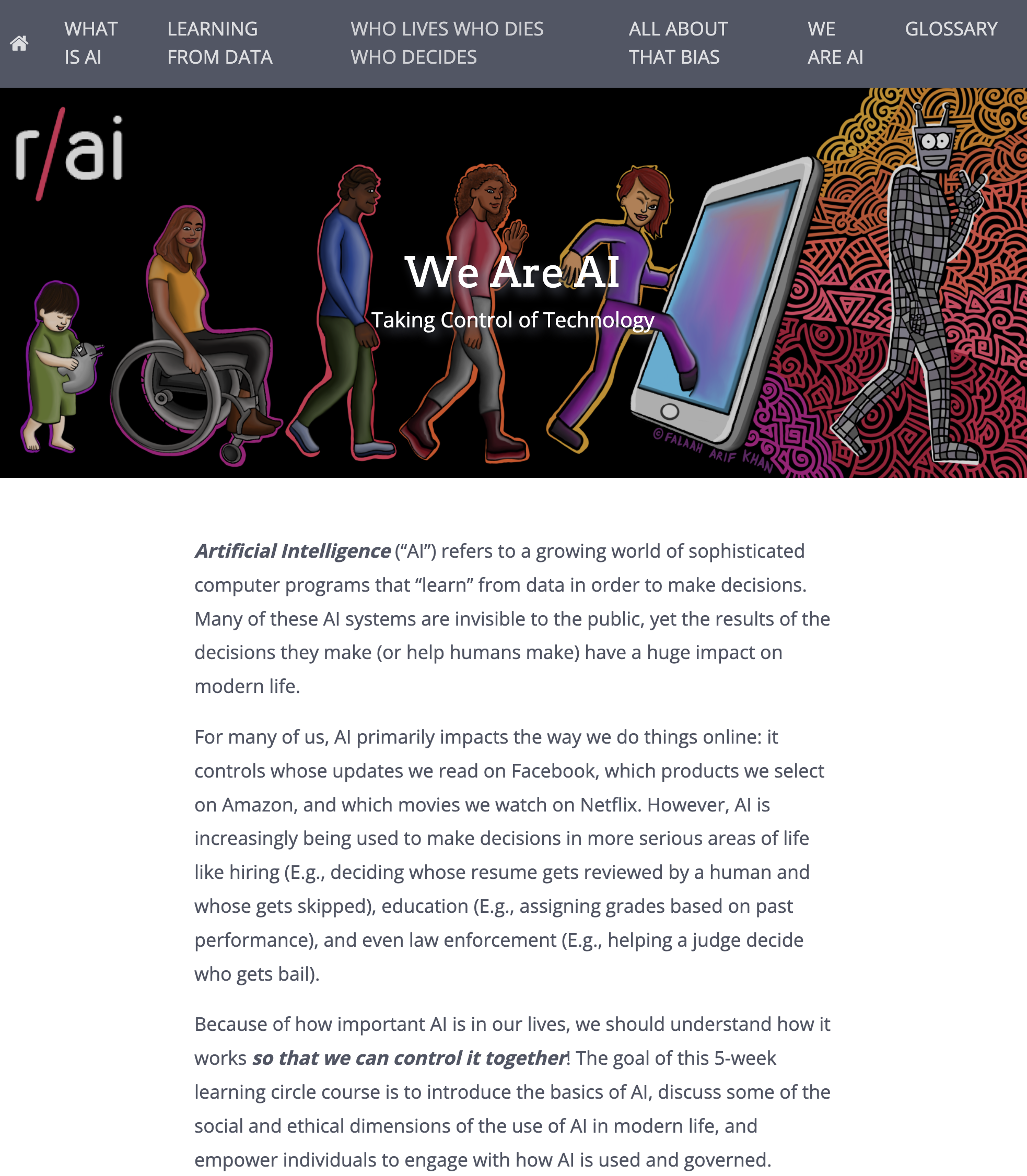}
\caption{The landing page of \emph{We Are AI: Taking Control of Technology}, stating the goals of the course and giving an outline of the modules.  The course website is publicly available at \url{https://r-ai.co/We-are-AI}.}
\label{fig:wai:landing}
\end{figure}

\emph{We Are AI} course materials are publicly available online.\footnote{\url{https://r-ai.co/We-are-AI}} Figure~\ref{fig:wai:landing} shows the landing page of the course website, featuring a verbal and visual overview of its goals. Notably, these materials include a comic book series of the same name---available in English, Spanish, and Ukrainian languages---that can be used independently.  This paper aims to introduce these resources to AI educators, researchers, and practitioners to support their public education efforts.

\section{Engaging the Public in Responsible AI}
\label{sec:rai}

Within research, several technical and design-based strategies exist to involve members of the public in the design, development, use, and oversight of AI.  One such strategy involves making black-box systems more interpretable via methods, design tenets, and visualizations, while balancing efficiency and accuracy goals  \cite{stoyanovich_nmi,mitchell_model_2019,gilpin_explaining_2018,ribeiro_why_2016,gebru_datasheets_2021,marcinkevics_interpretable_2023,chen_machine_2023}. The goal is to reduce perceived complexity and increase transparency of the inner workings of AI models.  Another approach involves soliciting multistakeholder participation throughout the AI creation process to better support those impacted by these systems. While participatory AI ideally seeks broader representation in AI development, close analysis reveals that stakeholders tend to be consulted marginally rather than empowered as decision-makers \cite{delgado_participatory_2023, liao_ai_2024}. This discrepancy stems from factors such as tight product release timelines, difficulties in integrating lay opinions into technical pipelines, and other prioritization issues. 

One approach to integrating diverse perspectives into AI has been the \emph{stakeholder-first perspective}, which emphasizes incorporating input from decision-makers, companies, shareholders, government and regulatory bodies, technologists, end users, and broader society at every stage of algorithmic system development \cite{bell_think_2023,gungor_creating_2020,lima_responsible_2020,miller_stakeholder_2022}.  Stakeholder-first works target AI practitioners (i.e., AI system designers, human-computer interaction specialists, engineers, and researchers) as principle change agents.  Fewer works integrate non-technical practitioners (community advocates, activists, policy makers, and organizational support staff) as full participants \cite{krafft_action-oriented_2021}.  A key challenge is the lack of accessible avenues for non-experts to understand the basic principles behind this technology, and use this knowledge to critique its applications and societal implications, and advocate for change.

\citet{dominguez_figaredo_responsible_2023} promote  AI literacy, enabling ordinary people to understand AI's societal implications, and present several case studies that advance this goal.  These include university courses that train the next generation of data scientists and engineers in technical and social best practices for RAI.  For example, the classroom-based Responsible Data Science course at New York University introduces undergraduate and graduate students to the basics of algorithmic fairness, the data science lifecycle, data protection, and transparency and interpretability~\cite{DBLP:journals/aiedu/LewisS22}.  During this semester-long course, students complete several thematic modules, in which content is delivered through case studies, fundamental algorithmic techniques, and hands-on exercises using open-source datasets and software libraries.\footnote{\url{https://r-ai.co/rds}}   While this and similar courses provide a valuable contribution to structured curricula, their reach is limited to a small group of students enrolled at specific institutions.  Publicly available courses remain scarce \cite{bell_atp, jagadish}, signaling the need for more accessible ways for diverse participants to learn about RAI.

\section{Teaching RAI in a Peer Learning Setting}
\label{sec:weareai}

The course \emph{We Are AI: Taking Control of Technology} comprises a series of open access modules that provide an overview of AI, explore its social and ethical implications, and empower individuals from diverse backgrounds to engage with its role in critical areas like hiring, education, and law enforcement.  Learners can benefit from the course irrespective of their technical or professional background, as it focuses on developing tools to critically and ethically engage with AI in a group setting.  \weareai is particularly valuable for people who are interested in learning about AI and its societal implications, but lack technical expertise required for traditional courses.

\subsection{Course Purpose}

The overarching idea behind this course is that technology and ethics are deeply intertwined, and people must take an active role in shaping how technology impacts them and others. To reflect this, the course interleaves technical, societal, and ethical concepts,  reinforcing the importance of human agency, as summarized in the course description:

\emph{[...] Because of how important AI is in our lives, we should understand how it works so that we can control it together! The goal of this 5-week learning circle course is to introduce the basics of AI, discuss some of the social and ethical dimensions of the use of AI in modern life, and empower individuals to engage with how AI is used and governed.''}

\subsection{Summary of Instructional Approach} 
In contrast to other publicly available courses on ethics and AI in an open format, \weareai adopts an intimate, scalable participatory model called a \emph{learning circle}, which has been shown to increase sustained learning and engagement with complex topics. A learning circle is a pedagogical model where small groups ``come together intentionally for the purpose of supporting each other in the process of learning''~\cite{collay_learning_1998}.  

There are two key qualities that characterize learning circle interactions.  The first is that participants join voluntarily, demonstrating their own intentionality rather than being driven by external pressures. The second is a shared purpose among participants; in this case, a collective commitment to learning about RAI.  This self-motivated aspect distinguishes our course from typical large open courses.

A learning circle is a peer-based modality without traditional teachers or students; instead, everyone learns the material together.  The group is guided by \emph{course facilitators}---individuals who, unlike instructors, are not required to have deep expertise in the topic, but are responsible for organizing the meetings and facilitating the discussions using prepared materials. Designating one or two facilitators at the start is essential to manage logistics and ensure the focus is placed on meaningful learning and collaboration.  The modules also contain helpful tips for facilitators and participants to scaffold the learning experience for all. 

A course participant can elect to become a facilitator in a future learning circle, naturally embedding the \emph{train-the-trainer} mechanism.  This dynamic is already evident in \weareai, where learners are stepping up to facilitate future sessions. In summary, by adopting the learning circles model for \weareai, we hope to offer a sustainable, scalable model for non-technical audiences to educate themselves and others about AI and RAI.   

\begin{figure}[t!]
\centering
\includegraphics[width=.45\textwidth]{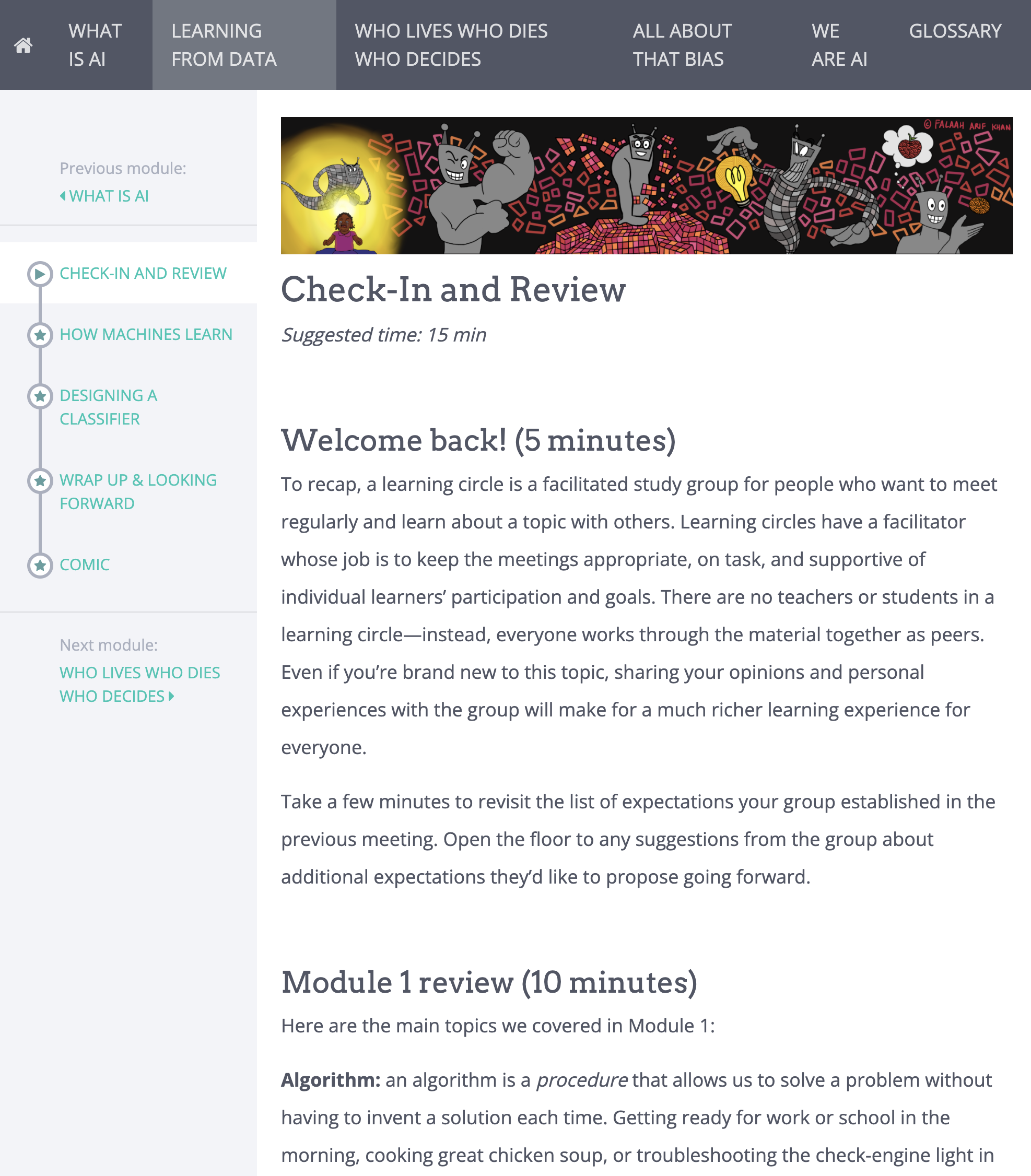}
\caption{A module of the We Are AI course.  All course modules follow a consistent structure.}
\label{fig:wai:module}
\end{figure}

\paragraph{Course timeline.} To date, \weareai has been offered in three iterations, both online and in person. 

The first iteration began in Spring 2021 through a collaboration between the NYU Center for Responsible AI (NYU R/AI), P2PU, an open-education non-profit, and the Queens Public Library in New York City. Course materials were developed and iteratively refined over several pilot runs, first with  librarians at our public library partner, facilitated by NYU R/AI instructors.  Later, two small groups of library patrons participated, one of which was facilitated by the librarians we had trained.

The second iteration, held in Summer 2023, was offered in person to 15 librarians and academic staff members at NYU. The third iteration, in Fall 2023, was delivered online to 26 professionals at NYU Abu Dhabi, one of the university's global locations.  Importantly, the second and third offerings allowed us to test the train-the-trainer model embedded in the course design. Two  librarians who participated in the Summer 2023 session successfully facilitated the course in Fall 2023.

All iterations of our course targeted librarians as the primary audience.  Librarians occupy a unique role as both knowledge brokers within their organizations and non-technical stakeholders with a deep interest in learning about the basics and implications of AI.  We will present qualitative and quantitative insights from the more recent full-scale offerings in a later section of this paper.

\subsection{Course Organization and Materials}
\label{sec:weareai:org}

All course materials, discussion prompts, and activities necessary to run a group with minimal preparation by the facilitator are incorporated into the course and available on the website. The intention is to enable anyone to use these resources to facilitate a learning circle or educate themselves about AI and RAI. No math, programming skills, or prior knowledge of AI is required.  

The course consists of five modules that progressively build foundational RAI knowledge. Participants meet synchronously for five 90-minute sessions (in person or online), with the group determining the meeting frequency. For example, the modules may be covered over 1-3 weeks or spread out to one module per week.  

The first session of a learning circle begins with  socialization activities to foster engagement. As is commonly done, in the \weareai learning circle, participants introduce themselves, share their motivations for participating, and discuss their general perceptions of AI. Before launching into the RAI material, the facilitator reviews the learning circle guidelines and sets group expectations.  These initial steps are important for encouraging openness and collaboration among participants.

Most sessions start with a recap of previous sessions, time for participants to raise questions or concerns, and review of any logistical matters. Participants then engage with the module content, which blends short videos, text, and collaborative activities.

\paragraph{Course modules.} 

The course website presents complete content for all five modules with a simple and consistent navigation structure. It also includes notes and tips for facilitators, along with a glossary.

Each module begins with a warm-up activity, followed by a brief instructional video (about 10 minutes). After the video, learners engage in discussion-based exercises in pairs or small groups, then share insights with the full group and discuss key takeaways.  All work is completed during the meeting, with no homework assigned---only optional supplemental readings. The modules are outlined below. 

\paragraph{}\emph{Module 1: What is AI?} Participants gain an overview of AI and explore algorithms: what they are, how they work, and how they relate to AI.  The instructional video explains algorithms, data, and decisions as the foundational components of an AI system. To build an intuitive understanding, participants consider the algorithm for baking bread.  The first interactive activity is writing down their own algorithm: a recipe for chicken noodle soup.  This is followed by a group discussion about whose recipe is ``the best'' (no one's!) and how our assessment of what's ``best'' depends on stakeholders (e.g., consumer, chef, cafe owner, parent), their values (e.g., taste, ease of preparation, nutritional value), and the context of use (e.g., home, cafe, school cafeteria).
    
\paragraph{}\emph{Module 2: Learning from data.} Participants explore how computers learn from data to make recommendations or assist in human decision-making, using relatable examples such as smart-home devices (e.g., how a smart light determines when to turn on or off) and credit card fraud detection. The module emphasizes the importance of applying the scientific method to assess prediction accuracy.  Participants come away with an understanding that how an algorithm functions, and the types of mistakes it tries to avoid, depend on the optimization goals defined by humans.

\begin{figure}[t!]
\centering
\includegraphics[width=.45\textwidth]{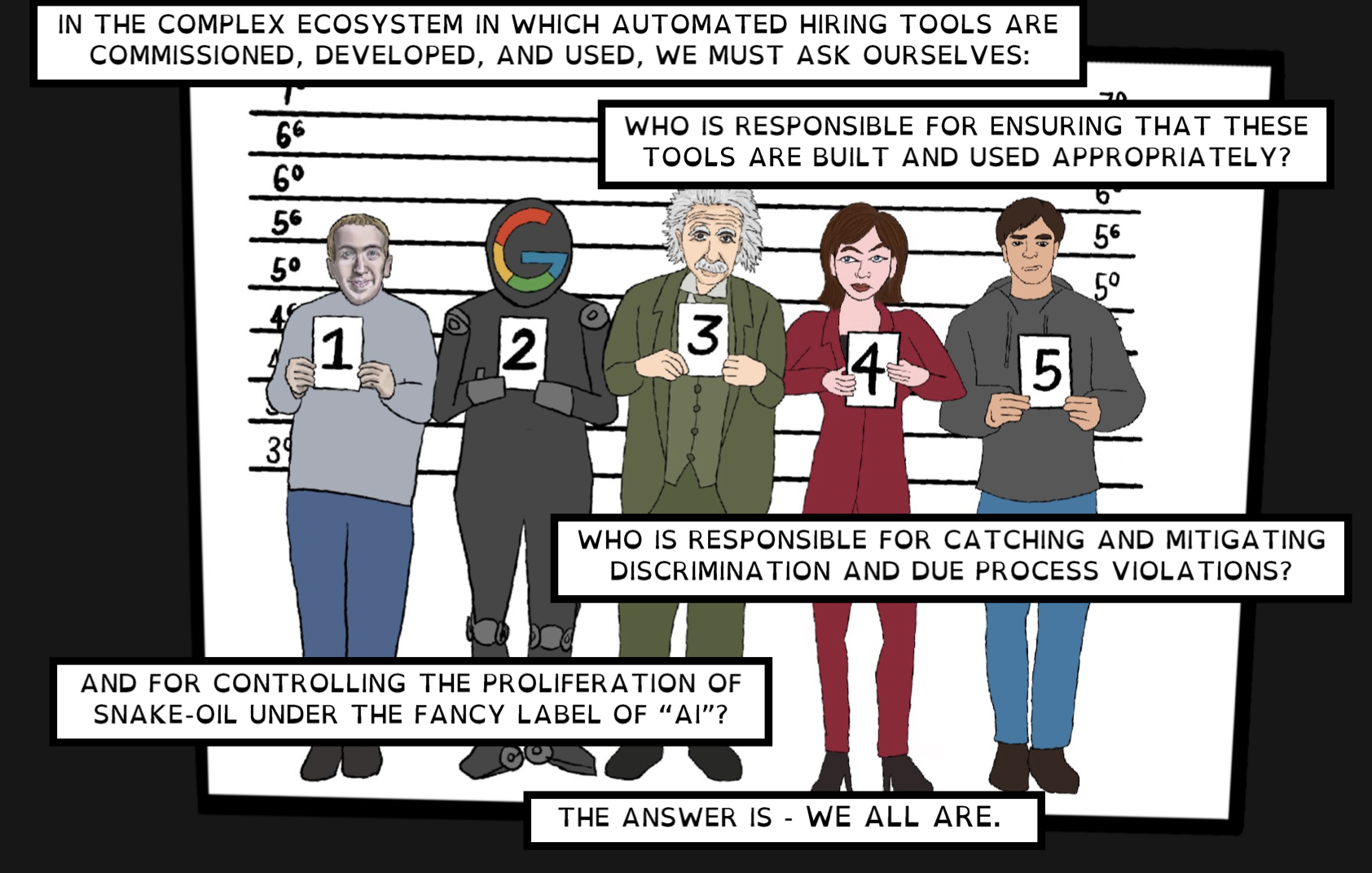}
\caption{A still from the We Are AI comic series.}
\label{fig:comics1:ads_comics}
\end{figure}

\paragraph{}\emph{Module 3: Who lives, who dies, who decides?} Participants examine key ethical issues surrounding AI and the need for active debate around AI and ethics.  The module begins with the classic trolley problem and critiques  \emph{algorithmic morality}---a purely utilitarian approach to AI ethics inspired by the problem.  Learners then engage in an extended discussion on the ethical challenges of autonomous vehicles\footnote{based on the story ``Who Killed Elaine Herzberg?'' at \url{https://onezero.medium.com/who-killed-elaine-herzberg-ea01fb14fc5e''}}, collaboratively completing an \emph{ethics matrix} for this  timely case study. 

\paragraph{}\emph{Module 4: All about that bias.} Participants analyze how decisions made with the help of AI are influenced by human bias.  After a warm-up discussion, a brief video introduces the concept of pre-existing, technical, and emergent bias, using the taxonomy of~\citet{DBLP:journals/tois/FriedmanN96}.  Learners are prompted to map this taxonomy to the context of algorithmic hiring systems.  They do this by interacting with a simulation of (highly discriminatory) hiring practices by a fictional company called ``Stock Mart,'' built using Teachable Machine~\footnote{\url{https://teachablemachine.withgoogle.com/}}, followed by a discussion. 

\paragraph{}\emph{Module 5: We are AI!} In this final module, learners reflect on their role in critically shaping the future of AI systems. This reflection is guided by a case study involving the use of automated decision systems (ADS) by a fictional (but realistic) Department of Child Protective Services to evaluate the risk of child death or injury in potential neglect and abuse cases.  This module also revisits key concepts from the course and invites participants to take action by learning more about AI and its impacts, teaching others, and stepping up to express opinions on AI-relevant laws and regulations.

\paragraph{Course website.} The course website is publicly available, requiring no sign-ups or registration to ensure broad accessibility.  Figure~\ref{fig:wai:module} shows a typical module, featuring a verbal and visual overview of session goals, facilitator guidance, and session recommendations. The course website also contains a glossary of terms and an extensive facilitator guide, which explains the course design rationale and provides specific content guidance.  We anticipate that the course content, glossary, and facilitator guide will continue to evolve as we refine the presentation and accessibility of the material. Additional modules will also be incorporated, as discussed in the final section of this paper.

\paragraph{Comic series.} We created the \weareai comic book series specifically as supplementary reading for the course, with one volume per module, see Figure~\ref{fig:comics1:ads_comics} for an example of a panel.  The comic is available for download in English, Spanish, and Ukrainian.\footnote{\url{https://r-ai.co/comics}} The design principles emphasize accessibility and ensure the course, including the comic, is suitable for a wide range of abilities and expertise levels.  The comic provides engaging content that can either guide sessions in real time or serve  as supplemental reading outside of class. We chose comics for the course because visual metaphors and humor effectively \emph{bridge the gap} between learners and the often intimidating topics of AI and ethics.

\section{Becoming Part of We Are AI: Participants' Experiences}
\label{sec:analysis}
As we already mentioned, we offered and facilitated \weareai at our academic institution, NYU, in Summer 2023.  The course attracted a diverse group of participants, including librarians and other staff supporting academic and research efforts. Two  librarians enthusiastically volunteered to facilitate the course in Fall 2023, which was specifically offered to librarians at NYU Abu Dhabi (see Table~\ref{tab:roles}).  

\begin{table}[b!]
\centering
\small 
\begin{tabular}{ l r r} 
 \textbf{Role} & \textbf{Summer 2023} & \textbf{Fall 2023} \\ 
 \hline
 administrator & 4 & 7 \\
 designer & 2 & 1 \\
 librarian  & 9 & 15 \\
 technologist  & 0 & 3 \\
 \hline
 \textbf{Total} &
 \textbf{15} &
 \textbf{26}\\
 \hline
\end{tabular}
\caption{Course participant roles.}
\label{tab:roles}
\end{table}

Since then, additional course participants have volunteered to facilitate new iterations of the course and have connected us with colleagues at other institutions interested in offering it to their librarians and professional staff. This provided initial evidence of the effectiveness of the train-the-trainer model embedded in the learning circles approach.

\subsection{Learning and Engagement around Ethical AI}
\begin{figure}[t!]
    \centering
    \includegraphics[width=\linewidth]{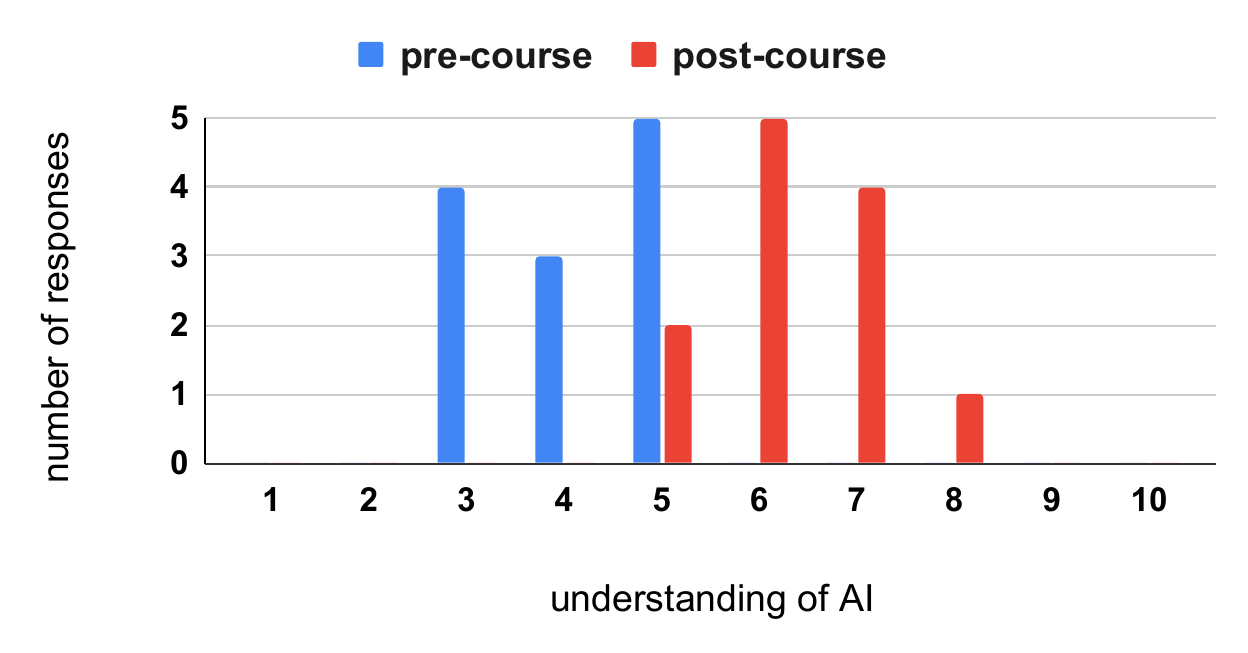}
    \caption{\emph{How do you rate your current understanding of AI?} \textbf{Summer 2023} pre-course and post-course ratings.}
    \label{fig:prepost:su23}
\end{figure}

\begin{figure}[t!]
    \centering
    \includegraphics[width=\linewidth]{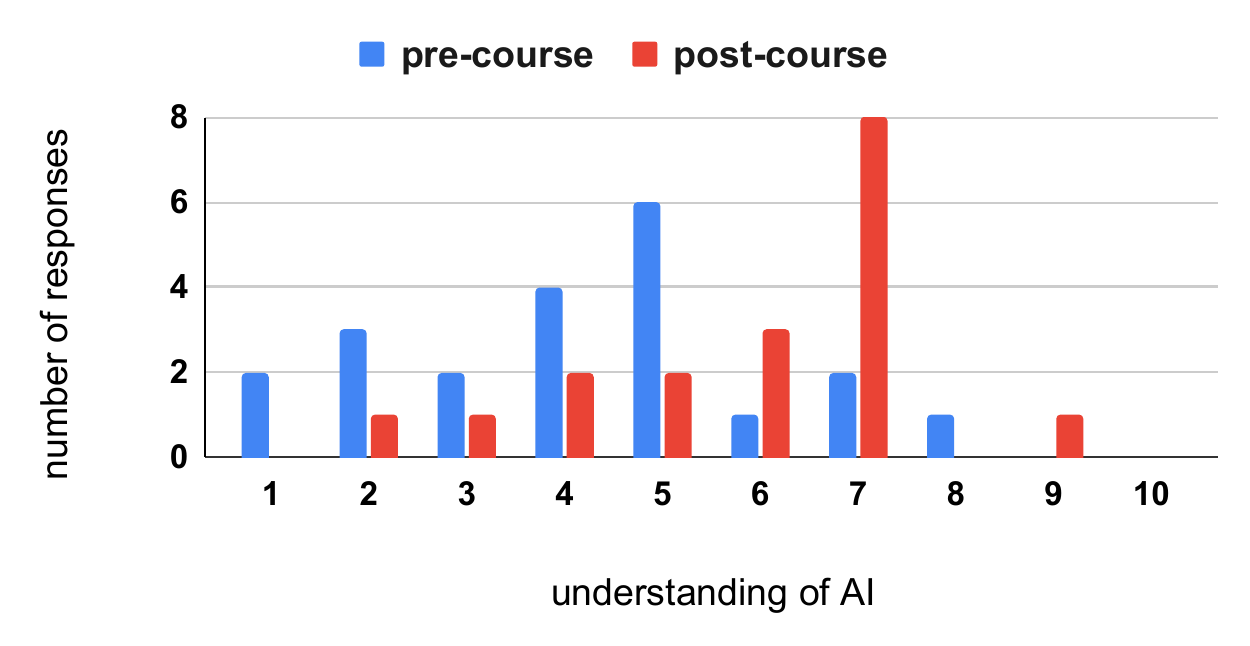}
    \caption{\emph{How do you rate your current understanding of AI?} \textbf{Fall 2023} pre-course and post-course ratings.}
    \label{fig:prepost:fa23}
\end{figure}

The primary goal of \weareai is to extend a foundational understanding of AI and RAI to non-technical audiences, including those directly impacted by these technologies. To evaluate learning outcomes, we conducted pre- and post-course self-assessment of RAI learning and general understanding, with results summarized in Figures~\ref{fig:prepost:su23} and~\ref{fig:prepost:fa23}.  Among the 15 participants in Summer 2023, 13 completed the pre-course survey, and 12 completed both surveys, resulting in a 92.3\% completion rate. For the 26 Fall 2023 participants, 21 completed the pre-course survey, and 18 completed both, yielding an 85.7\% completion rate between pre- and post-survey).  On average, participants rated their general understanding of RAI concepts and techniques significantly higher after completing the course than before: out of a maximum of 10, a weighted average of 4.08 and 4.19 for pre-course vs. 6.33 and 5.89 for post-course, in Summer 2023 and Fall 2023, respectively.  As one participant noted, ``I feel like I have a better base knowledge now with regards to how AI works and how folks are employing it.''
 
After the course, participants integrated key course concepts into their definitions of RAI.  They included phrases such as ``fairly impacting consumers,'' ``well thought out technology without harm to the public,'' ``biases are addressed,'' and ``AI systems are transparent, scientifically tested, and regulated''. Two-thirds of the respondents (66.7\%) reported that the course improved their understanding of AI substantially or very substantially, while all respondents indicated at least a moderate improvement. However, most participants still reported their understanding of algorithms as fuzzy or moderate, indicating an area for further course development.

\subsection{Empowering Individuals to Engage with AI}
An implicit goal of \weareai is to convey that everyone must play an active role in the design, development, use, and oversight of AI---we all are responsible. This idea is illustrated in Figure~\ref{fig:comics1:ads_comics} with a ``culpability line-up.''  As one participant noted, ``It was really significant that we are people first---We Are AI.''  Survey data indicates that participants' preconceived notions about AI included feelings of concern (Summer 2023) as well as curiosity and excitement (Fall 2023), which remained fairly consistent after completing the course.

Prior to the course, participants recognized both the benefits and the risks of using AI. Their engagement with AI reflected conflicting statements, such as ``I'm still cautious, but hopeful,'' and ``I think these tools are really interesting---but I feel the world is already relying on them too much.'' Several participants also expressed concerns about the rapid pace of the technological advancement, the lack of overt regulation, and unclear roles for individuals without technical expertise in AI development, use, and oversight. 

Throughout the course, participants developed two key strands of understanding.  The first centered on the importance of integrating more voices and perspectives into AI discussions. One participant summarized this with the quote, ``Responsible AI is the responsibility of everyone,'' emphasizing that the responsible use of AI is a societal matter, requiring collective participation. The course also highlighted how AI technologies are often deployed without proper consideration of their negative impacts, sparking a discussion on bias mitigation and inclusion.

The second strand of understanding focused on action steps participants felt prepared to take after learning the basics of Responsible AI.  Several participants cited gaining further technical expertise as a next action step. While the course is purposely designed to avoid technical explanations, some respondents indicated intentions to deepen their knowledge of machine learning systems---such as ``learn more about AI and how algorithms work''---now that they had a baseline understanding of the principles. 

Participants also intended to integrate lessons from the course into their professional and personal lives.  For example, one participant intended to limit the number of devices collecting their data, while another became ``more conscious about the way that [they] give out personal data.'' Others highlighted ``spreading the word, educating other people, and raising awareness about responsible AI.''  These learners expressed plans to launch their own learning circles and recruit novice learners, leveraging the train-the-trainer model. This potential proliferation of learning circles is a significant structural advantage of \weareai, enabling the course to scale with minimal resources.

\section{Lessons Learned}
\label{sec:lessons_rai} 

Throughout the project, we learned valuable lessons about what works well and what can be improved. Below, we summarize the key takeaways drawn from participant feedback, pre-and-post-course surveys, and observational notes. 

\paragraph{Scaffolded learning circles provide a framework for sustained and scalable collaborative learning.} While many existing courses on basic AI principles follow the massively open online course structure to guide large numbers of learners simultaneously, our model relies on the learning circles model.  This model typically involves smaller groups of no more than 30  participants who self-organize around a shared learning goal. Participants' collective motivation was echoed in their introductory remarks about why they elected to join the course.  The sessions alternate between small- and large-group discussions and hands-on activities, fostering dynamic exchanges among participants.

All participants expressed a desire to ``learn more and better understand how AI works'' and appreciated that ``everything was broken down and digestible.'' Given the  audience's varying levels of AI expertise, it was important to scaffold the experience to ensure that everyone benefited from participation.  Observations indicated that participants easily structured sessions, adhered to the agreed meeting schedule, and completed activities as structured on the website.  One participant remarked on ``the impact of the learning circle in a positive way. It is refreshing and surprising how well the learning circle works,'' while another noted the importance for a good learning experience. Consistent feedback indicated that participants learned AI basics and were ``much more confident'' about AI by the end of the course.

An additional benefit of learning circles is their integration of a train-the-trainer model, which implicitly prepares participants to become future facilitators based on topics of interest. Facilitators of the Fall 2023 course, who had been learners earlier that summer, remarked that the clear instructions effectively supported their own learning circles, particularly on topics covered in less depth, such as  techno-optimism, regulation, and mitigating negative effects of AI. This train-the-trainer model highlights the potential to scale \weareai while preserving the intimate structure of learning circles that fosters meaningful engagement.

\paragraph{Just-in-time engaging content encourages sustained learning. } Participants consistently found the comic series, AI examples, and illustrative case studies to be both engaging and appropriately targeted.  For example, participant noted the impact of the comic book's visual materials in reducing the intimidation factor: ``I really liked the illustrations. They made things understandable. AI can seem really over your head, but it brought it down to its brass tacks. It made it seem not so complicated.''  

Another participant underscored the importance of concrete case studies and examples: ``The case studies really opened up your mind to how to think about these things.''  Case Studies, a type of problem-based learning, are widely used across disciplines and linked to improved knowledge transfer, real-world problem-solving, and recognition of intersecting values and impacts of complex phenomena ~\cite{hackney_using_2003, wassermann_getting_1993}.  \citet{kreber_learning_2001} argues that case studies provide essential experiential learning opportunities, fostering reflection,  reconceptualization, and applied use of  knowledge. We developed high-quality case studies that capture actual AI systems, platforms, and scenarios, prompting participants to reconcile tensions resulting from disparate stakeholder impacts.  The case studies are accessible to all, regardless of technical expertise, as they are grounded in real-world examples. Several participants appreciated the ability to revisit these materials on the website and share them with colleagues.

\paragraph{Need to clarify and expand technical scope.} One critical piece of feedback highlighted the need to communicate the course's goals and content more explicitly and accurately. Though the learning outcomes and themes are explicit within the presented materials, participants suggested emphasizing that \weareai (1) is an introductory course on AI as how it is deployed in society, (2) is not intended as a technical course, and (3) does not currently focus on generative AI. We linked the feedback on generative AI to the widespread media coverage of tools such as ChatGPT, which often leads to  conflation of AI with generative AI. In designing the course, we deliberately selected examples that show that the expectations and context of AI usage often matter more than the technology's sophistication. That said, we plan to address this gap by developing a new module and two case studies involving ChatGPT for an upcoming offering of the course. See Next Steps for further discussion.

\paragraph{Need for additional scaffolding to streamline course facilitation.} While we made efforts to provide structure to the course flow and allow flexibility based on group dynamics, feedback  indicated a need for more scaffolding. This is particularly important given the complexity and evolving nature of AI, as well as the need to introduce both conceptual information and high-level technical principles. One planned addition is to provide recommendations for effective facilitators, emphasizing skills such as managing schedules, keeping time, and guiding the group through the session outline.  Additionally, we intend to provide more examples to help facilitators foster a comfortable learning environment and navigate  conflicts that may arise during heated breakout or group discussions.

\paragraph{Need for contextual engagement.}  One takeaway was the need to design materials with greater intentionally around context. Some participants noted the importance of ``giving more context to the video'' content, a point we are addressing by providing short primers to set expectations.  This corresponds with learning research that emphasizes the value of establishing concrete expectations prior to presenting material.  Related to contextualizing individual content, it is important to acknowledge the context in which the materials were created. We wrote the examples and case studies from a U.S. perspective, and offering this course to an international audience revealed the need to customize the scenarios.  For instance, child protective services, an agency supporting children at risk of neglect and abuse, may not resonate across all cultural contexts.  Instead of rewriting our scenarios completely, we plan to integrate prompts that encourage participants to adapt the lessons to their personal, national, and cultural contexts, and to reflect on how these lessons can be integrated into their specific roles and responsibilities.

Another contextual consideration is the need to promote more diverse discussions.  Several participants noted that breakout sessions often involved the same people. A possible solution is for facilitators to incorporate randomized breakout groups online and promote mixing groups during in-person sessions.

\paragraph{Need to anticipate and support next stages of learning and action.} A major theme that emerged was participants' curiosity about ``what’s next'' after completing the course.  Feedback indicated that participants felt their understanding of AI had increased and that they were motivated to continue learning.  Some planned to enroll in open courses on machine learning, while others planned to facilitate their own learning circles.  One suggested improvement was to create a second part of the course, tailored for those with advanced beginner or intermediate knowledge. One participant asked: ``What happens when you know a little bit more?   How do we increase [our] level of expertise?''  A follow- up course could delve deeper into topics such as AI regulation and bias mitigation strategies while introducing relevant technical concepts.  Several learners also reflected on broader legal structures, with one noting,  ``The course made me think about how we can play a role in regulation. I hadn’t thought much about regulation.''  

Another suggested improvement for the course is to provide participants with ways they can act to improve AI for society, such as ``how everyday people can get involved in actions that support the ethical use of AI.'' Several participants mentioned the need for ``concrete'' and ``actionable'' steps to take.  To address this, we propose adding a module focused on opportunities for ordinary citizens to engage in the design, development, and use of AI.  In the United States, such involvement could begin at the local or state level.  
\section{Conclusion and Next Steps}
\label{sec:next}

In this paper we presented \emph{We Are AI: Taking Control of Technology}, an open course designed to foster  engaged learning around responsible AI. The course follows a learning circle structure, where participants with a shared interest in the topic come together to learn and connect.  We hope that this self-motivated approach will inspire others in the community to collaborate in building a deeper understanding of the pedagogical needs of Responsible AI and to develop and share much-needed concrete educational materials and methodologies. 

Much work remains to address the educational and training needs of current and future data scientists, decision-makers who use AI, policy makers, auditors, and the public at large. Additionally, there is a need to rigorously evaluate the effectiveness of educational methodologies for these audiences. Both  directions are integral to our ongoing work.

We are seeing increasing demand from colleagues at academic institutions and community organizations, both in the U.S. and internationally, to offer \weareai to their audiences. To meet this need, we plan to add modules tailored to specific learners, prioritizing librarians, academic administrators, and current and future education professionals. emerging, we are developing a module on AI-based chatbots, drawing examples from recent case studies to highlight both the opportunities and challenges of responsibly deploying  generative AI in critical domains, such as K-12 education \footnote{\url{https://www.nytimes.com/2023/09/11/learning/would-you-want-an-ai-tutor.html}} or in giving (often incorrect) legal advice to small and medium-sized businesses\footnote{\url{https://themarkup.org/news/2024/03/29/nycs-ai-chatbot-tells-businesses-to-break-the-law}}.  

\begin{links}
     \link{Course website}{https://r-ai.co/We-are-AI}
      \link{Comic books}{https://r-ai.co/comics}
\end{links}

\section{Acknowledgements}
\label{sec:ack}

This work would not have been possible without collaboration and input from many of my colleagues. Dr. Mona Sloane, Meghan McDermott, Becky Margraf, Grif Peterson, Jeffrey Lambert, Sadie Coughlin-Prego, and Kaven Vohra all contributed to the development and refinement of the materials and methodologies used in this course.  

This research was supported in part by NSF Awards No. 1922658, 2326193, and 2312930.

\bibliography{rds}

\end{document}